\documentclass{article}

\usepackage{PRIMEarxiv}

\usepackage[utf8]{inputenc} 
\usepackage[T1]{fontenc}    
\usepackage{hyperref}       
\usepackage{url}            
\usepackage{booktabs}       
\usepackage{amsfonts}       
\usepackage{nicefrac}       
\usepackage{microtype}      
\usepackage{lipsum}
\usepackage{fancyhdr}       
\usepackage{graphicx}       
\graphicspath{{media/}}     

\usepackage{times}
\usepackage{latexsym}

\usepackage[T1]{fontenc}

\usepackage[utf8]{inputenc}

\usepackage{microtype}

\usepackage{inconsolata}
\usepackage{microtype}
\usepackage{graphicx}

\usepackage{hyperref}

\usepackage[textsize=tiny]{todonotes}
\usepackage{amssymb}

\usepackage{natbib} 
\usepackage{adjustbox}
\usepackage{graphicx}
\usepackage{caption}
\usepackage{subcaption}

\usepackage{amsmath}
\usepackage{bm}
\usepackage{multirow}
\usepackage{xcolor}
\usepackage[utf8]{inputenc}
\usepackage{mathtools}
\usepackage{bbm}
\usepackage{physics}
\usepackage{float}
\usepackage{booktabs}
 \usepackage{array, makecell}
 \usepackage{algorithm}
 \usepackage{algorithmic}
\usepackage{tabularx}
\usepackage{amsthm}

\usepackage{hyperref}
\title{Prospective Evaluation of Multimodal Respiratory Failure Prediction: Do Chest X-Rays Improve Performance Beyond EHR Signals?
}

\author{
 Xiaolei Lu, Shamim Nemati
}

\begin{document}
\maketitle

\begin{abstract}
Early prediction of respiratory failure is critical for timely clinical intervention in intensive care units. Existing electronic health record (EHR)-based models can continuously monitor physiologic deterioration, but they may not fully capture pulmonary pathophysiology reflected in chest radiographs (CXRs). In this study, we ask whether CXR information improves prospective prediction of invasive mechanical ventilation beyond EHR signals alone. We develop a gated multimodal framework that integrates structured EHR time-series data with CXR foundation-model representations. The gating module adaptively controls the contribution of imaging features based on patient-specific clinical context, allowing the model to selectively rely on imaging information when it is informative. We prospectively evaluate the framework for predicting invasive mechanical ventilation within 24 hours in ICU patients and compare it with an established EHR-only model (Vent.io), physician predictions obtained at matched clinical time points, and alternative multimodal variants. The gated multimodal models achieved higher discrimination than the EHR-only baseline, with AUROC values of 0.860 and 0.858 using REMEDIS and MedInsight CXR representations, respectively, compared with 0.752 for Vent.io. Relative to physician predictions, the multimodal framework substantially improved sensitivity while maintaining favorable specificity. Compared with the EHR-only model, multimodal integration increased specificity and positive predictive value, suggesting that CXR information can refine risk estimation in selected patients. These findings support adaptive multimodal fusion as a practical strategy for incorporating imaging into prospective respiratory failure prediction.
\end{abstract}


\section{Introduction}

Respiratory failure is a major cause of clinical deterioration in critically ill patients and frequently necessitates invasive mechanical ventilation (IMV). Delayed recognition of impending respiratory decompensation may result in missed opportunities for timely intervention, leading to prolonged intensive care unit (ICU) stay, increased complications, and worse clinical outcomes. Early identification of patients at elevated risk may enable clinicians to initiate interventions such as high-flow nasal cannula (HFNC), non-invasive ventilation (NIV), antibiotics, corticosteroids, fluid management, or escalation of monitoring before severe deterioration occurs\citep{grotberg2023management, fan2018acute}.

Machine learning approaches have demonstrated promising capability for predicting respiratory failure using electronic health record (EHR) data\citep{shashikumar2021development, lam2024development}. By continuously incorporating physiological measurements, laboratory values, medications, comorbidities, and temporal trends, these models provide real-time estimates of future respiratory risk. Prior studies have shown that EHR-based systems can predict respiratory deterioration hours before the onset of invasive mechanical ventilation and may provide clinically actionable warning signals in both retrospective and prospective settings.

However, structured EHR variables may incompletely characterize pulmonary disease processes. Important aspects of respiratory pathology, including pulmonary edema, infiltrates, consolidation, pleural effusions, and progression of radiographic abnormalities, may not be fully represented through continuously monitored physiologic variables alone. Chest radiographs (CXRs) provide complementary information regarding lung pathology and remain among the most frequently obtained imaging studies in ICU patients. Clinicians often rely on imaging findings to understand mechanisms underlying respiratory deterioration and to guide treatment decisions.

Despite the widespread use of CXRs in clinical practice, an important question remains unresolved:

\begin{center}
\textit{Do chest radiographs provide clinically meaningful predictive information beyond continuously monitored EHR signals?}
\end{center}

This question is important because imaging information may not contribute uniformly across all patients and time points. In some cases, structured EHR signals may already capture a patient's physiologic state sufficiently well. In other situations, radiographic evidence of worsening pulmonary disease may substantially refine risk estimation. Therefore, simply incorporating imaging features into predictive models does not necessarily guarantee improved performance.

Existing multimodal prediction approaches commonly employ static fusion strategies, such as direct feature concatenation or late-stage fusion, which implicitly assume that different modalities contribute similarly across patients \citep{lee2025multi, tandon2023hybrid}. However, the value of imaging information is highly dependent on clinical context. For example, chest radiographs may provide limited additional information in clinically stable patients but may become highly informative in patients with rapidly evolving pulmonary conditions. Consequently, a fixed multimodal integration strategy may fail to capture the variable importance of imaging information across different patient states.

Beyond evaluating multimodal integration itself, another important challenge is understanding how machine predictions compare with physician judgment. Physicians routinely synthesize physiologic trends, radiographic findings, and contextual clinical information when assessing respiratory deterioration. However, these assessments can vary across providers and are difficult to quantify systematically. Comparing model predictions with physician assessments in prospective clinical workflows provides an opportunity to determine whether multimodal systems capture complementary information and whether they improve identification of patients who eventually deteriorate.

To address these challenges, we propose a gated multimodal framework for prospective respiratory failure prediction that adaptively integrates structured EHR signals and chest radiograph representations. The framework consists of an EHR encoder, a CXR image encoder based on pretrained foundation models, and a gating module that dynamically estimates the contribution of imaging information according to patient-specific clinical context. Rather than assuming imaging contributes equally across all patients, the proposed framework selectively adjusts the influence of radiographic information during prediction.

We perform prospective evaluation for prediction of invasive mechanical ventilation within the subsequent 24 hours in ICU patients and compare the proposed framework against an established EHR-based model (Vent.io), physician predictions collected at matched clinical time points, and alternative multimodal variants using different CXR foundation representations. Our objective is not only to evaluate whether imaging improves overall discrimination, but also to understand how multimodal integration changes clinically relevant operating characteristics including sensitivity, specificity, and positive predictive value.

The main contributions of this study are summarized as follows:

\begin{itemize}

\item We develop a gated multimodal framework that adaptively integrates structured EHR signals and chest radiograph foundation-model representations for respiratory failure prediction.

\item We prospectively evaluate whether chest radiographs provide incremental predictive value beyond an established EHR-based model and physician assessments obtained at matched clinical time points.

\item We characterize how multimodal integration influences clinically relevant operating characteristics, including sensitivity, specificity, and positive predictive value.

\end{itemize}

\section{Methodology}

The proposed framework consists of three components:
(1) an EHR encoder,
(2) an image encoder, and
(3) a gated multimodal integration module for respiratory failure prediction.
\subsection{EHR Encoder}

We employ an EHR encoder based on the Vent.io\citep{lam2024development} architecture to learn patient representations from structured clinical data. At each hourly timestamp, the input consists of both static and dynamic variables, including demographics, comorbidities, medications, vital signs, laboratory measurements, and derived temporal features.

To account for irregular sampling in clinical observations, we incorporate a time-since-last-measurement (TSLM) mechanism. For each dynamic variable $x_j$, we additionally compute its measurement recency $\Delta t_j$, representing the elapsed time since the feature was last observed. A lightweight TSLM layer applies a learnable decay function to reduce the contribution of stale measurements while incorporating recency information into the feature representation.

Formally, given structured input at time $t$:

\begin{equation}
X_e=
\{X_s, X_d, \Delta t\}
\end{equation}

where $X_s$ denotes static features, $X_d$ represents time-varying clinical variables, and $\Delta t$ corresponds to feature-wise TSLM values, the transformed features are concatenated and passed into the EHR encoder:

\begin{equation}
h_e=f_e(X_e;\theta_e)
\end{equation}

where $h_e\in\mathbb{R}^{d}$ represents the learned latent clinical representation of the patient's current physiological state.

The EHR encoder follows the Vent.io backbone architecture, which combines TSLM-enhanced feature processing with a multilayer perceptron (MLP) encoder for prospective respiratory failure prediction.

\subsection{Image Encoder}

Chest radiographs are processed using pretrained chest X-ray foundation models to extract imaging representations. Unlike task-specific models trained solely for respiratory failure prediction, foundation models provide rich visual features learned from large-scale medical imaging data and capture diverse radiographic patterns associated with pulmonary disease.

Given an input chest radiograph $X_c$, the image encoder produces an imaging representation:

\begin{equation}
z_c=f_c(X_c;\theta_c)
\end{equation}

where $z_c\in\mathbb{R}^{d_c}$ denotes the feature embedding generated by the pretrained foundation model.

Since the EHR and imaging modalities originate from different feature spaces, the extracted imaging features are passed through a projection network to map them into a shared latent space before multimodal fusion:

\begin{equation}
h_c=
g_p(z_c;\theta_p)
\end{equation}

where $h_c\in\mathbb{R}^{d}$ represents the transformed imaging representation aligned with the EHR latent space, and $g_p(\cdot)$ denotes a learnable multilayer projection network.

In our experiments, we investigate alternative image encoders using different chest radiograph foundation models, including REMEDIS \citep{azizi2023robust} and MedInsight \citep{neupane2024medinsightmultisourcecontextaugmentation} representations, to evaluate how pretrained imaging representations influence multimodal respiratory failure prediction.

\subsection{Gated Multimodal Integration}

Rather than directly concatenating representations from both modalities, we dynamically estimate the contribution of imaging information using a gating mechanism.

The gate value is computed as:

\begin{equation}
g=
\sigma
(
W[h_e;h_c]+b
)
\end{equation}

where

\begin{equation}
g\in [0,1]
\end{equation}

represents the estimated importance of chest radiograph information.

The final multimodal representation is computed as:

\begin{equation}
h=
(1-g)h_e
+
gh_c
\end{equation}

where smaller gate values indicate stronger reliance on EHR signals and larger values indicate increased influence from imaging information.

The integrated representation is subsequently passed into a prediction layer to estimate the probability of invasive mechanical ventilation within the next 24 hours.

\section{Experiment}

\subsection{Dataset}
We developed and evaluated the model using de-identified ICU encounters from adult patients ($\geq$18 years) admitted to UC San Diego Health (UCSDH). Encounters between January 1, 2016 and December 31, 2023 were used for model development and retrospective validation. For the multimodal analysis, encounters without matched chest radiographs (CXR) within the predefined temporal matching window were excluded. The final multimodal development cohort consisted of 17,167 encounters, the test cohort consisted of 4,292 encounters. Additional internal retrospective test performance is presented in Appendix \ref{in}.Institutional Review Board approval was obtained under protocol \#800258 (``A Real-Time Multimodal Data Integration Model for Prediction of Respiratory Failure in Patients with COVID-19'').

We performed prospective evaluation using ICU patients from two medical intensive care units across two University of California San Diego (UCSD) hospitals. The study was conducted from October 2024 to May 2025 under an approved institutional review board protocol. Table~\ref{tab:patient_characteristics} summarizes the demographic and clinical characteristics of the study cohort. Detailed preprocessing procedures and EHR feature descriptions are provided in Appendix \ref{data}.

To align imaging with structured EHR observations, chest radiographs (CXRs) were matched to EHR samples at corresponding prediction timestamps. Within the same patient encounter, the most recent available CXR representation was forward-filled to subsequent EHR samples until a newer radiograph became available. For ICU encounters without CXR acquisition during the ICU stay, we additionally considered images obtained from other clinical departments within the preceding 72 hours. Samples without corresponding CXR information after this matching process were excluded from multimodal analysis. 

To ensure fair comparison across prediction approaches, model predictions, physician assessments, and Vent.io outputs were generated at matched timestamps using only contemporaneously available clinical information.

\subsection{Implementation Details}

The EHR encoder was initialized using pretrained Vent.io parameters because of its previously demonstrated performance in prospective respiratory failure prediction. During multimodal training, the entire framework, including the Vent.io backbone, image projection module, and gated fusion network, was jointly optimized in an end-to-end manner.

Model optimization was performed using the Adam optimizer. Hyperparameters associated with the multimodal framework, including learning rate, batch size, hidden dimensions, regularization parameters, and dropout rates, were selected through Bayesian hyperparameter optimization on the training set.

To facilitate comparison across methods, operating thresholds were selected according to a predefined sensitivity target. Specifically, thresholds were determined on the validation set to achieve approximately 60\% sensitivity and subsequently fixed during test evaluation.
\begin{table}[t]
\centering
\caption{Baseline characteristics of ventilated and nonventilated encounters.}
\label{tab:patient_characteristics}
\begin{tabular}{lcc}
\hline
\textbf{Characteristic} & \textbf{Nonventilated} & \textbf{Ventilated} \\
\hline

\# Encounters (\%) & 233 (94.3\%) & 14 (5.7\%) \\

Age, years & 60.4 [49.5--74.0] & 64.7 [59.5--75.6] \\

Male, n (\%) & 120 (51.5\%) & 6 (42.9\%) \\

\hline
 \multicolumn{3}{l}{\textbf{Race, n (\%)}} \\
\hline

\hspace{3mm} White & 130 (55.8\%) & 6 (42.9\%) \\

\hspace{3mm} Black or African American & 13 (5.6\%) & 1 (7.1\%) \\

\hspace{3mm} Asian & 17 (7.3\%) & 2 (14.3\%) \\

\hspace{3mm} Other race or mixed race & 68 (29.2\%) & 3 (21.4\%) \\

\hline
\multicolumn{3}{l}{\textbf{Clinical severity}} \\
\hline

\hspace{3mm} Charlson Comorbidity Index & 2.0 [1.0--3.0] & 2.5 [1.0--3.0] \\

\hspace{3mm} SOFA Score & 5.0 [3.0--7.0] & 9.0 [4.5--12.0] \\

\hspace{3mm} Vasopressor administration, n (\%) & 157 (67.4\%) & 13 (92.9\%) \\

\hline
\end{tabular}
\end{table}

\subsection{Physician Performance}

A total of 247 physician assessments were collected during the prospective evaluation cohort, including 233 nonventilated encounters (94.3\%) and 14 ventilated encounters (5.7\%). Physicians predicted impending intubation within the subsequent 24 hours in 12 cases (4.9\%), while predicting no need for intubation in 235 cases (95.1\%).

Physicians correctly identified 3 of the 14 patients who subsequently required invasive mechanical ventilation, corresponding to a sensitivity of 0.214. In contrast, physicians correctly identified 224 of the 233 patients who did not require intubation, yielding a specificity of 0.961. Overall, physician prediction achieved a balanced accuracy of 0.588.

Despite the relatively low sensitivity, physicians reported high confidence in their assessments. Confidence scores, measured on a scale from 0 (not at all confident) to 10 (extremely confident), demonstrated a median value of 8.0 and a mean of 7.4 $\pm$ 2.3.

\subsection{Model Performance}

The gated multimodal framework integrating EHR and chest radiograph (CXR) information demonstrated improved overall discriminative performance compared with the EHR-only baseline model (Vent.io), achieving an AUROC of 0.860 using REMEDIS CXR foundation representations, compared with 0.752 for the EHR-only model.

Using the predefined operating threshold, the gated multimodal model with REMEDIS image representations achieved a specificity of 0.831, sensitivity of 0.391, and positive predictive value (PPV) of 0.243. In comparison, the EHR-only model achieved a specificity of 0.368, sensitivity of 0.739, and PPV of 0.139.

Using the same gated multimodal framework but replacing the CXR representation with MedInsight foundation embeddings resulted in comparable overall discrimination (AUROC 0.858), while further increasing specificity to 0.910 and PPV to 0.318. However, this improvement was accompanied by a reduction in sensitivity to 0.304.

Overall, incorporating CXR information within the gated multimodal framework improved specificity and precision relative to the EHR-only baseline while maintaining strong overall discrimination performance. Although both REMEDIS and MedInsight representations yielded similar AUROC values, they demonstrated different operating characteristics, suggesting that the choice of foundation representation influences the sensitivity-specificity tradeoff of the multimodal framework.
\begin{table}[t]
\centering
\caption{Performance comparison between physician predictions and predictive models for identifying intubation within the subsequent 24 hours.}
\label{tab:model_comparison}

\begin{tabular}{lcccc}
\toprule
\textbf{Predictor} & \textbf{AUC} & \textbf{Specificity} & \textbf{Sensitivity} & \textbf{PPV} \\
\midrule

Physician
& --
& 0.961
& 0.214
& 0.250 \\

Vent.io (EHR only)
& 0.752
& 0.368
& 0.739
& 0.139 \\

Gated Multimodal (REMEDIS)
& 0.860
& 0.831
& 0.391
& 0.243 \\

Gated Multimodal (MedInsight)
& 0.858
& 0.910
& 0.304
& 0.318 \\

\bottomrule
\end{tabular}

\vspace{2mm}

\footnotesize{
AUC = area under the receiver operating characteristic curve; PPV = positive predictive value. Physician predictions are binary and therefore do not have a corresponding AUC value.
}

\end{table}
\section{Discussion}

The primary objective of this study was to investigate whether chest radiographs provide additional predictive value beyond continuously monitored EHR signals for prospective respiratory failure prediction. Our results suggest that incorporating imaging information can improve prediction performance, although the benefit is not reflected as a uniform increase across all evaluation metrics. Compared with the EHR-only baseline model, multimodal integration substantially increased specificity and positive predictive value while maintaining strong overall discrimination performance. These findings suggest that CXR information may refine EHR-derived risk estimates by providing complementary pathophysiologic information that is not fully represented by structured clinical variables alone.

An important observation is that multimodal integration altered model operating characteristics rather than simply increasing predictive accuracy. The EHR-only model demonstrated higher sensitivity but lower specificity, whereas the multimodal approaches shifted predictions toward greater specificity and precision. One possible interpretation is that imaging information provides additional evidence that reduces uncertainty in patients with elevated EHR-derived risk, thereby decreasing false-positive predictions. Clinically, this behavior may be valuable in reducing unnecessary alerts or interventions in ICU settings where alarm burden remains a significant challenge.

Comparison with physician assessments provides additional insights into the potential role of multimodal prediction systems. Physicians achieved high specificity but relatively low sensitivity, indicating a tendency toward conservative prediction behavior in which only patients perceived to be at high risk were identified. In contrast, the predictive models identified substantially more patients who eventually required intubation while maintaining reasonable specificity. This difference likely reflects distinct decision-making processes. Physician assessments may incorporate broader contextual reasoning and clinical judgment, whereas prediction models systematically integrate large numbers of physiologic and imaging variables that may be difficult to synthesize continuously in real time. Rather than replacing clinician judgment, these findings suggest that multimodal prediction systems may serve as complementary tools for identifying patients at risk of deterioration earlier than may otherwise be recognized.

Interestingly, the two multimodal approaches using REMEDIS and MedInsight foundation representations demonstrated different prediction behaviors despite achieving similar overall discrimination performance. While both approaches improved specificity relative to the EHR-only baseline, the MedInsight-based model further increased specificity and positive predictive value at the expense of sensitivity. One possible explanation is that the distinct pretraining strategies and learned image representations of these foundation models emphasize different radiographic characteristics. Consequently, the imaging representations may influence not only overall predictive performance but also how risk information is prioritized within the multimodal framework.

This study has several limitations. First, the prospective cohort was collected from a single healthcare system and may not fully represent broader patient populations or institutional workflows. Second, CXR acquisition was not performed uniformly across patients, introducing potential selection bias in multimodal samples. Third, physician assessments were collected at individual prediction time points and may not fully capture longitudinal clinical reasoning processes. Future work should investigate multicenter validation, more comprehensive physician-model interaction studies, and methods for improving interpretability of multimodal predictions. In particular, understanding when imaging contributes meaningful information and identifying clinically relevant radiographic representations may further improve prospective respiratory failure prediction systems.

\nocite{langley00}

\bibliographystyle{unsrtnat}
\bibliography{reference}  

@article{fan2018acute,
  title={Acute respiratory distress syndrome: advances in diagnosis and treatment},
  author={Fan, Eddy and Brodie, Daniel and Slutsky, Arthur S},
  journal={Jama},
  volume={319},
  number={7},
  pages={698--710},
  year={2018},
  publisher={American Medical Association}
}

@article{grotberg2023management,
  title={Management of severe acute respiratory distress syndrome: a primer},
  author={Grotberg, John C and Reynolds, Daniel and Kraft, Bryan D},
  journal={Critical Care},
  volume={27},
  number={1},
  pages={289},
  year={2023},
  publisher={Springer}
}

@article{lam2024development,
  title={Development, deployment, and continuous monitoring of a machine learning model to predict respiratory failure in critically ill patients},
  author={Lam, Jonathan Y and Lu, Xiaolei and Shashikumar, Supreeth P and Lee, Ye Sel and Miller, Michael and Pour, Hayden and Boussina, Aaron E and Pearce, Alex K and Malhotra, Atul and Nemati, Shamim},
  journal={JAMIA open},
  volume={7},
  number={4},
  pages={ooae141},
  year={2024},
  publisher={Oxford University Press}
}

@article{shashikumar2021development,
  title={Development and prospective validation of a deep learning algorithm for predicting need for mechanical ventilation},
  author={Shashikumar, Supreeth P and Wardi, Gabriel and Paul, Paulina and Carlile, Morgan and Brenner, Laura N and Hibbert, Kathryn A and North, Crystal M and Mukerji, Shibani S and Robbins, Gregory K and Shao, Yu-Ping and others},
  journal={Chest},
  volume={159},
  number={6},
  pages={2264--2273},
  year={2021},
  publisher={Elsevier}
}

@inproceedings{lee2025multi,
  title={Multi-VentNet: Multimodal Fusion Framework for Predicting Need for Mechanical Ventilation},
  author={Lee, Ye Sel and Lu, Xiaolei and Miller, Michael and Lam, Jonathan Y and Shashikumar, Supreeth and Nemati, Shamim},
  booktitle={2025 IEEE 22nd International Symposium on Biomedical Imaging (ISBI)},
  pages={1--5},
  year={2025},
  organization={IEEE}
}

@article{tandon2023hybrid,
  title={A hybrid decision tree and deep learning approach combining medical imaging and electronic medical records to predict intubation among hospitalized patients with COVID-19: algorithm development and validation},
  author={Tandon, Pranai and Ghanavati, Sahar and Cheetirala, Satya Narayana and Timsina, Prem and Freeman, Robert and Reich, David and Levin, Matthew A and Mazumdar, Madhu and Fayad, Zahi A and Kia, Arash and others},
  journal={JMIR formative research},
  volume={7},
  number={1},
  pages={e46905},
  year={2023},
  publisher={JMIR Publications Inc., Toronto, Canada}
}

@article{azizi2023robust,
  title={Robust and data-efficient generalization of self-supervised machine learning for diagnostic imaging},
  author={Azizi, Shekoofeh and Culp, Laura and Freyberg, Jan and Mustafa, Basil and Baur, Sebastien and Kornblith, Simon and Chen, Ting and Tomasev, Nenad and Mitrovi{\'c}, Jovana and Strachan, Patricia and others},
  journal={Nature Biomedical Engineering},
  volume={7},
  number={6},
  pages={756--779},
  year={2023},
  publisher={Nature Publishing Group UK London}
}

@misc{neupane2024medinsightmultisourcecontextaugmentation,
      title={MedInsight: A Multi-Source Context Augmentation Framework for Generating Patient-Centric Medical Responses using Large Language Models}, 
      author={Subash Neupane and Shaswata Mitra and Sudip Mittal and Noorbakhsh Amiri Golilarz and Shahram Rahimi and Amin Amirlatifi},
      year={2024},
      eprint={2403.08607},
      archivePrefix={arXiv},
      primaryClass={cs.CL},
      url={https://arxiv.org/abs/2403.08607}, 
}

\medskip


\appendix

\section{Dataset, Model and Baselines}
\subsection{Dataset}
\label{data}
\noindent{\textbf{Patient inclusion and exclusion criteria}}. Patients were included in the respiratory failure prediction analysis if they had an ICU stay of at least five hours, were not mechanically ventilated before ICU admission, and had documented vital signs and laboratory values prior to the prediction start time. Those with a Do Not Resuscitate (DNR) order were excluded, and data within 24 hours of surgery were omitted to avoid bias from surgery-related ventilation. Monitoring continued until mechanical ventilation was initiated or ICU discharge. To ensure sufficient data, predictions began four hours post-admission and were updated hourly using the latest clinical information.

\noindent{\textbf{Data abstraction and processing}}. We extracted EHR data encompassing 50 vital signs and laboratory measurements, 6 demographic features, 12 Systemic Inflammatory Response Syndrome (SIRS) and Sequential Organ Failure Assessment (SOFA) criteria, 12 medication categories, and 62 comorbidities. To handle varying sampling frequencies, vital signs and laboratory values were aggregated into hourly time-series bins, with multiple measurements per hour summarized using the median. Data updates occurred hourly, with missing values carried forward for up to 24 hours if no new data were available. Remaining missing values were imputed using the mean. Additionally, we derived 150 features from the 50 vital signs and laboratory measurements, including baseline values (mean over the previous 72 hours), local trends (change since the last measurement), and time since last measured (TSLM).

\noindent{\textbf{Clinical labeling scheme for the various physiological states of respiratory failure}}. Table \ref{tab} lists the labeling criteria used in \citep{lam2024development}.  

\begin{table}
\centering
\caption{Criteria of clinical labeling scheme.}
\begin{tabular}{|l|l|c|}
\hline
\textbf{Condition}           & \textbf{Criteria}                                                                 & \textbf{Points} \\ \hline
\multirow{5}{*}{PaO\textsubscript{2}/FiO\textsubscript{2} (not NaN)} & \(200 < \text{PaO}_2/\text{FiO}_2 \leq 300 \, \text{mmHg}\)                           & 1              \\ \cline{2-3} 
                              & \(\text{PaO}_2/\text{FiO}_2 \leq 200 \, \text{mmHg}\) (severe hypoxemia)           & 2              \\ \cline{2-3} 
                              & IMV \(\leq 24 \, \text{hours}\)                                                   & 3              \\ \cline{2-3} 
                              & \(\text{PaO}_2/\text{FiO}_2 \leq 200 \, \text{mmHg}\) and IMV \(\leq 24 \, \text{hours}\) & 4              \\ \cline{2-3} 
                              & IMV \(> 24 \, \text{hours}\)                                                      & 5              \\ \hline
\multirow{5}{*}{SpO\textsubscript{2}/FiO\textsubscript{2} (not NaN)} & \(141 < \text{SpO}_2/\text{FiO}_2 \leq 221 \, \text{mmHg}\)                          & 1              \\ \cline{2-3} 
                              & \(\text{SpO}_2/\text{FiO}_2 \leq 141 \, \text{mmHg}\) (severe hypoxemia)           & 2              \\ \cline{2-3} 
                              & IMV \(\leq 24 \, \text{hours}\)                                                   & 3              \\ \cline{2-3} 
                              & \(\text{SpO}_2/\text{FiO}_2 \leq 141 \, \text{mmHg}\) and IMV \(\leq 24 \, \text{hours}\) & 4              \\ \cline{2-3} 
                              & IMV \(> 24 \, \text{hours}\)                                                      & 5              \\ \hline
\end{tabular}
\label{tab}
\end{table}

\noindent{\textbf{Encounter-level evaluation}}. Table \ref{tab:eval_metrics} lists the details of evaluation metrics.

\begin{table}[h!]
\caption{Definitions of evaluation metrics based on predictions and labels.}
\centering
\begin{tabular}{@{}ll@{}}
\toprule
\textbf{Metric}            & \textbf{Definition} \\ \midrule
\textbf{True Positive (TP)} & A positive prediction (\( \text{predictions[t]} \geq \text{threshold} \)) where there is at least \\ &  one
positive label within the prediction window (up to 24 hours before \( T_0\textsuperscript{a} \)). \\ \midrule
\textbf{False Positive (FP)} & A positive prediction (\( \text{predictions[t]} \geq \text{threshold} \)) where no positive labels\\
                             & exist within the prediction window (up to 24 hours before \( T_0 \)). \\ \midrule
\textbf{False Negative (FN)} & A negative prediction (\( \text{predictions[t]} < \text{threshold} \)) where a positive label \\
                             & exists within the prediction window (up to 24 hours before \( T_0 \)). \\ \midrule
\textbf{True Negative (TN)}  & A negative prediction (\( \text{predictions[t]} < \text{threshold} \)) where no positive labels \\
                             & exist throughout the evaluated timestamps. \\ \bottomrule
\end{tabular}
\label{tab:eval_metrics}
\vspace{0.2em}
\par \textsuperscript{a} \( T_0 \) is defined as the first timestamp where a patient is ventilated based on the simultaneous recording of PEEP and FiO$_2$.
\end{table}

\section{Internal Test Performance}
\label{in}

We evaluated unimodal and multimodal architectures on the internal test cohort. To assess the contribution of each modality, we compared EHR-only and chest radiograph (CXR)-only models trained on the same cohort. We further evaluated two multimodal fusion strategies: (1) an attention-based fusion architecture (Fusion 1) adapted from prior multimodal learning frameworks \citep{lee2025multi}, and (2) our proposed gated multimodal fusion architecture (Fusion 2).

The EHR-only model achieved an AUROC of 75.70, outperforming the CXR-only model, which achieved an AUROC of 71.10. Incorporating imaging information through multimodal fusion improved predictive performance over either unimodal approach. The attention-based fusion architecture achieved an AUROC of 77.61, while the proposed gated fusion architecture demonstrated the best overall performance with an AUROC of 83.33.

To further evaluate the effect of image foundation models, we additionally tested a ResNet-based CXR encoder (cxr-50x1-remedis-m) and a transformer-based MedImageInsight encoder. Across both settings, the gated multimodal architecture consistently outperformed unimodal baselines and the attention-based fusion approach, suggesting that adaptive gating mechanisms may better capture complementary information between structured EHR signals and chest radiographs.
\begin{figure}[t]
    \centering
    \includegraphics[width=\columnwidth]{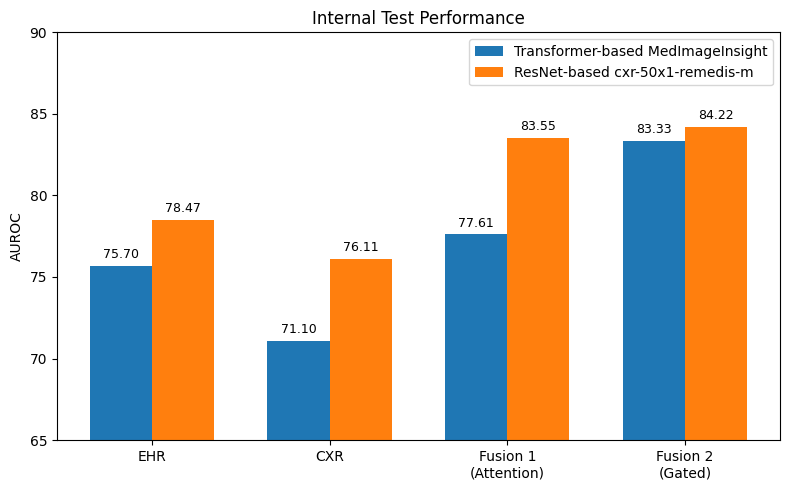}
    \caption{
    Internal prospective test performance across unimodal and multimodal architectures using different chest radiograph (CXR) foundation models. 
    EHR denotes the structured EHR-only model, while CXR denotes the image-only model trained on the same cohort. 
    Attention Fusion corresponds to the attention-based multimodal fusion architecture, whereas Gated Fusion corresponds to the proposed gated multimodal fusion architecture. 
    Results are shown using both the transformer-based MedImageInsight encoder and the ResNet-based cxr-50x1-remedis-m encoder.
    }
    \label{fig:internal_test}
\end{figure}

\end{document}